\documentclass[ usenatbib]{mnras}
\usepackage{times,graphicx,amssymb, amsmath, natbib}
\usepackage[T1]{fontenc}

\begin{document}

\title[Constraining Neutrino mass using the large scale HI distribution in the Post-reionization epoch]
{Constraining Neutrino mass using the large scale HI distribution in the Post-reionization epoch} 

\author[Pal,  \& Guha Sarkar ]
{Ashis Kumar  Pal\thanks{E-mail: ashispal475@gmail.com}, 
 Tapomoy Guha Sarkar\thanks{E-mail: tapomoy1@gmail.com} \\
Department of Physics, Birla Institute of Technology and Science, Pilani, Rajasthan, 333031. India.
}

\maketitle

\date{\today}

\begin{abstract}
The neutral intergalactic medium in the post reionization epoch allows
us to study cosmological structure formation through the observation
of the redshifted 21 cm signal and the Lyman-alpha forest. We
investigate the possibility of measuring the total neutrino mass
through the suppression of power in the matter power spectrum. We
investigate the possibility of measuring the neutrino mass through its
imprint on the cross-correlation power spectrum of the 21-cm signal
and the Lyman-alpha forest. We consider a radio-interferometric
measurement of the 21 cm signal with a SKA1-mid like radio telescope
and a BOSS like Lyman-alpha forest survey. A Fisher matrix analysis
shows that at the fiducial redshift $z = 2.5$, a 10,000 hrs 21-cm
observation distributed equally over 25 radio pointings and a
Lyman-alpha forest survey with 30 quasars lines of sights in $1 {\rm
  deg}^2$, allows us to measure $\Omega_{\nu}$ at a $3.25 \%$ level.
A total of 25,000 hrs radio-interferometric observation distributed
equally over 25 radio pointings and a Lyman-alpha survey with $\bar n
= 60 {\rm deg}^{-2}$ will allow $\Omega_{\nu}$ to be measured at a $
2.26 \%$ level.  This corresponds to an idealized  measurement of $\sum m_{\nu}$
at the precision of $(100 \pm 2.26) \rm meV$ and $f_{\nu}=
\Omega_{\nu}/ \Omega_{m}$ at $2.49 \%$ level.

\end{abstract}


\def\n{{\bf{\hat{n}}}}
\def\be{\begin{equation}}
\def\ee{\end{equation}}
\def\bear{\begin{eqnarray}}
\def\ear{\end{eqnarray}}
\def\nline{\nonumber \\}
\def\f{\frac}
\def\de{{\rm d}}
\def\del{\partial}

\def\R{{\cal R}}
\def\etar{\eta_{\rm LSS}}
\def\valpha{\boldsymbol{\alpha}}
\def\Fcal{\mathcal{F}}
\def\begm{\begin{pmatrix}}
\def\enm{\end{pmatrix}}
\def\eps{\varepsilon}
\def\l{\ell}
\def\r{r}

\def\matrixsymbol{\sf}

\def\apj{ApJ}
\def\apjl{ApJL}
\def\mnras{MNRAS}
\def\prd{PRD}
\def\k{{\bf k}} 
\def\r{{\bf r}}
\def\HI{{\rm HI}}
\def\x{\vec{x}}
\def\s{\tilde{r}}

\section{Introduction}
It has been established from neutrino oscillation experiments that
neutrinos have mass. There are
profound implications of these particles being massive and
cosmological observations are expected to put strong constrains on
neutrino physics. Several cosmological probes are used in this
regard \citep{lesgourgues2013neutrino}. We believe that at least two of the three species of neutrinos
are non-relativistic today. Neutrinos are in expected to be in thermal
equilibrium with CMBR in the early Universe and their number is
fixed. Accordingly their fractional contribution to the present day
matter budget is given by $ f_{\nu} = \frac{\Omega_{\nu}}{\Omega_{m}}
$ where $ \Omega_{\nu} $=$\sum\nolimits_{i} m_{i}/93.14 h^{2} $eV
\citep{neutrino11, neu22} with $m_i$ denoting the mass of each neutrino
species.  When the temperature of the universe is very high they can
be treated as a part of radiation and after the CMB temperature drops
below their masses, they can contribute to the matter density of the
Universe.  The free streaming behaviour of neutrinos causes them to
wipe out fluctuations on scales smaller than the horizon scale
when the
neutrinos became non-relativistic \citep{ hu_mdm, 1999ApJ...511....5E, lesgourgues2013neutrino}. The free-streaming comoving vector
is given by $k_{fs} = \sqrt{3/2} H(z)/ v (1 + z)$ where $v$ denotes
the free-streaming speed. When neutrinos are relativistic, this
coincides with the horizon scale, and when neutrinos become non-relativistic at $ z = z_{nr} $ \citep{k_nr} it becomes
\[k_{nr} \sim \sqrt{\frac{3}{2}} \frac{H(z_{nr})m_{\nu}}{ 3 T_{\nu}(z_{nr}) (1 +
z_{nr})}. \]  For modes  $ k < k_{nr}$
neutrinos behave like usual dark matter and there is no suppression of
power. Modes with $k> k_{fs}$ have neutrino perturbations wiped out and
consequently CDM power spectrum is also suppressed.
Cosmological measurement of neutrino mass \citep{ly_neutrinomass1, hannes2003,
  neuPritchard1, cosparam2, sudeep1, 2015JCAP02045P, valentino, relicnu14, allison15,chen15} depends directly on the level of
precision at which this suppression of power in the matter power
spectrum can be detected. Whereas CMBR gives stringent constraints on
neutrino mass \citep{ CMBlensing-neutrino, neucmbb, neucmb1, planck13, planck15} it is necessary to obtain
measurements from the low redshift Universe. Matter power spectrum for
$z < 1$ is highly non-linear and neutrino mass measurements have
degeneracies with competition from dark energy \citep{hannes2005}.  

Mapping the neutral hydrogen distribution (HI) in the
post-reionization epoch using the observation of the redshifted 21 cm
signal towards measurement of the matter power spectrum has been
studied as a means to measure neutrino mass \citep{mcquinn,
  neuPritchard1, neucmbb, Shimabukuro14, 21CMB, Oyama15}.  At
redshifts $z < 6$, HI responsible for the 21 cm signal lies
predominantly in the Damped Lyman Alpha (DLA) clouds \citep{wolfe05}
and the diffuse collective 21 cm emission from these clouds form a
background in radio observations. The redshifted 21-cm diffuse
emission from the post-reionization epoch is well modelled using a
constant neutral fraction \citep{xhibar1, xhibar2}, and a bias
function \citep{bagla2, tgs2011}. The diffuse neutral gas distribution
in the same redshift range can also be probed using the distinct
absorption features of the Lyman-alpha forest \citep{rauch98}.  Large
astrophysical foregrounds \citep{ fg4, fg1, fg5, fg9, ghosh2011,
  alonso2014} come in the way of detecting the 21 cm signal.  The
cross correlation of the 21 cm signal with other cosmological probes
like the Lyman-alpha forest and Lyman-Break galaxies, has been
proposed as a viable way \citep{tgs5, TGS15, navarro2} to mitigate the
effect of foregrounds. The flux through the Lyman-alpha forest and the
redshifted 21 cm signal can both be modeled as biased tracers of the
underlying dark matter distribution and are expected to be correlated
\citep{mcd03, bagla2, tgs2011, navarro}. The cross-correlation signal
is a direct probe of the matter power spectrum over a large redshift
range in the post reionization epoch.

In this paper we investigate the constraints on total neutrino mass
using the cross correlation of the Lyman-alpha forest and the 21-cm
signal from the post reionization epoch. The suppression of power on
scales $k > k_{nr}$ allows us to constrain the neutrino mass.  In
  this paper we constrain the parameters $\Omega_m$ and $\Omega_{\nu}$
  using the cross-correlation signal. These are the only two free
  parameters in our analysis.  The values of $\Omega_{\Lambda} = 0.685, \Omega_b h^2 = 0.02222, h =0.6731, n_s=0.9655, \sigma_8 = 0.829$  are fixed to their fiducial values as obtained from (\citet{planck15}Table 4: TT + lowP) for this analysis.
$\Omega_{K}$ varies but is not treated as a  free independent parameter.  We
  consider a future radio interferometric observation of the 21 cm
  using a telescope like
  SKA1-mid \footnote{https://www.skatelescope.org/} and a
  BOSS \footnote{https://www.sdss3.org/surveys/boss.php} like
  Lyman-alpha forest survey for obtaining predictions for bounds on
  the parameters.

\section{Cross correlation signal and estimation of Neutrino mass}

The cross-correlation of the redshifted 21-cm signal and the
transmitted flux through the Lyman-alpha forest has been proposed and
studied as a cosmological probe of structure formation and background
evolution in the post-reionization epoch \citep{TGS15}. The two signals of
interest owe their origins to neutral hydrogen (HI) in distinct
astrophysical systems at redshifts $ z \leq 6$ but are expected to be
correlated on large scales.

Bulk of the HI in the post-reionization epoch is believed to be housed
in the dense self-shielded DLA systems. These
clumps are the dominant source of the 21-cm emission in this
epoch. The signal from individual DLA sources is rather weak and
resolving individual sources is also rather unfeasible. The collective
emission from the clouds, however form a diffuse background in all
radio observations at observing frequencies less than $1420$ MHz. Tomographic
imaging using the low resolution intensity mapping of this diffuse background
radiation is a potentially rich probe of cosmology \citep{ poreion1, poreion4, param2, wyithe08, param3, poreion0, camera13, cosmo14}.

Along a line of sight ${\bf \hat{n}}$ traversing a HI cloud at
redshift $z$ the CMBR brightness temperature changes from $T_{\gamma}$
to $T(\tau_{_{21}})$ owing to the emission/absorption corresponding to
the spin flip Hyperfine transition at $\nu_e = 1420$ MHz in the rest
frame of the gas.  \be T(\tau_{_{21}}) = T_{\gamma} e^{-\tau_{_{21}}}
+ T_s ( 1 - e^{-\tau_{_{21}}} ) \ee where $\tau_{_{21}}$ is the 21-cm
optical depth and $T_s$ is the spin temperature \citep{hirev1}.  The quantity of
interest in radio observations at a frequency $\nu = \nu_e/(1+z)$ is
the excess brightness temperature $T_b ( {\bf\hat{n}}, z)$ redshifted
to the observer at present. This is given by \be T_b( {\bf\hat{n}}, z)
= \frac{T(\tau_{_{21}}) - T_{\gamma}}{1 + z} \approx \frac{(T_s -
  T_{\gamma})\tau_{_{21}}}{1 + z}.\ee Writing $\textbf{r} \equiv (r
{\bf\hat{n}}, z)$, where $r$ is the comoving distance corresponding to
$z$, we have the fluctuations in $T_b( {\bf\hat{n}}, z)$ given by 
   $\delta_T(\textbf{r}) = \bar T(z) \times \eta_{\rm
  HI}(\textbf{r})$ \citep{bharad04}  , where
\begin{equation}
\bar T(z)=4.0 {\rm mK} (1+z)^{2}\left(\frac{\Omega_{b0}h^{2}}{0.02}\right)\left(\frac{0.7}{h}\right) \left( \frac{H_{0}}{H(z)}\right)
\end{equation}
and
\begin{equation}
\begin{split}
\eta_{HI}( r {\bf\hat{n}}, z)=\bar{x}_{HI}(z) \left\{ \left ( 1 - \frac{T_\gamma}{T_s}\right)\left[\delta_{H}(z,{\bf\hat{n}})-\frac{1+z}{H(z)}\frac{\partial v}{\partial r}\right] \right. \\ 
\left.+
 \frac{T_\gamma}{T_s} s \delta_{H}({\bf\hat{n}}r,z) \right\}
\end{split}
\end{equation}
Here $ \bar{x}_{HI}(z) $ denote the mean neutral fraction, $
\delta_{H}(z,{\bf\hat{n}}) $ denotes the HI density fluctuations and
$s$ is a function that takes into account the fluctuations of the spin
temperature assuming it to be related to the HI fluctuations. The peculiar velocity of the gas, $v$ introduces the  anisotropic term $ (1+z)/ H(z)\frac{\partial v}{\partial r}$.

In the post reionization epoch one has $T_{\gamma}/T_s << 1 $ whereby  the 21 cm signal is seen in emission and we have
\be 
\eta_{_{HI}}( r {\bf\hat{n}}, z)=\bar{x}_{_{HI}}(z)\left[\delta_{H}(z,{\bf\hat{n}})-\frac{1+z}{H(z)}\frac{\partial v}{\partial r} \right].
\ee
 In Fourier space the  fluctuation in 21-cm excess brightness temperature  $\delta_{ T} (\textbf{r})$ is denoted by 
$\Delta_T (\textbf{k})$ and is given by 
\begin{equation}
\Delta_{T}(\textbf{k})=C_{T}[1+\beta_{T}\mu^{2}]\Delta(\textbf{k})
\end{equation}
where $\Delta(\textbf{k})$ is the Fourier transform of the  dark matter over density $\delta$  and  we have assumed that the peculiar velocity is sourced by dark matter over density. The quantity $\beta_T$ is similar to the redshift space distortion \citep{hamilton} parameter and $ \mu = {\bf\hat{n}} \cdot  {\bf\hat{k}}$. 
The redshift dependent quantity $C_T$ is given by 
$C_T = \bar T(z) \bar{x}_{HI}(z) b_T$ where 
$b_T$ denotes a bias that relates the  HI
fluctuations in Fourier space  $\Delta_H(\textbf{k})$  to the underlying dark matter
fluctuations $\Delta(\textbf{k})$ as 
$\Delta_H (\textbf{k})  =  b_T \Delta (\textbf{k})$.
The 21 cm bias has been studied in several independent studies \citep{ bagla, marin, tgs2011}.
The small scale fluctuations clearly have a scale dependent bias which grows monotonically with $k$.
There is also some additional scale dependence owing to fluctuations in the ionizing background \citep{poreion0}.
Numerical simulations show that on large scales the bias is  however found to be a constant increasing only with redshift \citep{ bagla, marin, tgs2011}.
We adopt a linear bias model \citep{tgs2011} in this work.
It is important to note that in cosmologies with massive neutrinos HI is more clustered. This is an additional effect which occurs because we fundamentally believe that the gas is contained in halos and halos are rarer in models with neutrino mass and are thereby more biased \citep{massara}. This enhances the 21-cm power spectrum by a roughly scale independent factor of $1.15$ at $ z = 2.5$ \citep{navarro-viel}. 

 Whereas the dense clumpy HI sources the 21 cm emission, the diffuse  intergalactic HI in the post
reionization epoch also produces distinct absorption lines in the
spectra of background quasars namely the Lyman-alpha forest \citep{reion, bolton}. The 21 cm signal and the Lyman-alpha forest arises from two distinct
astrophysical systems. Bulk of the 21 cm emission comes from the dense
 HI  clouds which are self shielded
from the ionizing photons. The Lyman-alpha forest, on the other hand
consists of absorption lines in the Quasar spectra which originates
from tiny fluctuations in the low density diffuse HI along the line of
sight.  The Flux $ \mathcal{F}$ through the Lyman-alpha forest is
modeled using the fluctuating Gunn-Peterson approximation  \citep{gunnpeter, bidav} \be
\frac{\Fcal}{\bar{ \Fcal}} = {\rm exp}^{- A( 1 + \delta)^\alpha } \ee
where $\alpha$ is related to the slope of the temperature density
relationship $\gamma$ as $\alpha = 2 - 0.7 (\gamma - 1 )$, and $A$ \citep{bolton}  is
a redshift dependent parameter depending on several astrophysical and
cosmological parameters like the photo-ionization rate, IGM temperature
and the evolution history \citep {pspec, pspec1}. The relationship between the observed flux
through the Lyman-alpha forest and the underlying dark matter
distribution is non-linear\citep{slosar2}. However, on large scales the smoothed
Lyman-alpha forest flux traces the dark matter fluctuations \citep{vielmat, saitta, slosar1} via a
bias. The precise analytic understanding about the bias is not very robust and has been studied under certain approximations.
 However,  a linear bias model is largely supported by numerical
simulations of the Lyman-alpha forest \citep{mcd03}.

The fluctuations of the flux through the Lyman-alpha forest
$\Delta_{\Fcal}$ can hence be related to the dark matter fluctuations
in Fourier space as \be \Delta_{\Fcal} (\textbf{k}) = C_{\Fcal} [ 1 +
  \beta_{\Fcal} \mu^2 ] \Delta (\textbf{k}) \ee The parameters $(
C_{\Fcal}, \beta_{\Fcal})$ are independent of one another and depend
on $A$, $\gamma$ and the flux probability distribution function.  The
predictions for $\beta_{\Fcal}$ from particle-mesh simulations
indicate that increasing the smoothing length has the effect of
lowering the value of $\beta_{\Fcal}$ which is ideally set by choosing
the smoothing scale at the Jean's length which is in turn sensitive to
the IGM temperature. We adopt the values $(C_{\Fcal}, \beta_{\Fcal}) = ( -0.15, 1.11) $ from simulations of Lyman-alpha forest \citep{mcd03}.

Being  tracers of the underlying dark matter distribution 
the Lyman-alpha forest and 21 cm signal are expected to be correlated.
The cross correlation of the Lyman-alpha forest and the 21cm signal has been 
proposed as a cosmological probe \citep{ tgs5, TGS15}. 
Apart from mitigating the serious issue of 21 cm foregrounds, the cross correlation  if detected, ascertains the cosmological origin of the signal as opposed to the auto correlation 21 cm signal. Several advantages of the cross-correlation 
have been studied in earlier works \citep{TGS15}.
The  3D cross correlation power spectrum is defined through 
\be
\langle \Delta_{\Fcal}(\textbf{k}) \Delta_{T}^{\ast}(\textbf{k}^{\prime})\rangle=(2\pi)^{3} \delta^{3}(\textbf{k}-\textbf{k}^{\prime})P_{\Fcal T}(\textbf{k}),
\ee  with 
\be
P_{\Fcal T}(\textbf{k})=C_{\Fcal} C_{T}(1+\beta_{\Fcal}\mu^{2})(1+\beta_{T}\mu^{2})P(\textbf{k}).
\ee
where $P(\textbf{k})$ is the dark matter power spectrum.

Observationally the Lyman-alpha forest survey as well as radio
observations of the 21 cm signal will probe certain volumes of the
Universe. The cross-correlation can however be computed only in the
overlapping region denoted by ${\cal V}$. 
 The Lyman-alpha surveys typically shall cover a larger volume and thus ${\cal V}$ is set by the 21-cm observations. If $B$ be the observational bandwidth of the 21 cm observation and if $\theta_a \times \theta_a$ is the angular patch that the radio telescope probes,  then 
 $ {\cal V} = B \dfrac{dr}{d\nu} \times r^2 \theta_a^2 $.

The observed 21 cm signal denoted by $ \Delta_{TO}$  shall include a noise term 
and may be written as 
\be
\Delta_{T0}(\textbf{k})=\Delta_{T}(\textbf{k})+\Delta_{NT}(\textbf{k})
\ee
Similarly the observed Lyman-alpha forest transmitted flux fluctuation in Fourier space 
may be written as

\be
\Delta_{\mathcal{F}o}(\textbf{k})= \int \tilde{\rho}(\textbf{k} -\textbf{K} ) \Delta_{\mathcal{F}}(\textbf{K}) ~d^3\textbf{K}  +\Delta_{N\mathcal{F}}(\textbf{k})
\ee
which involves a convolution of the three dimensional field with a normalized sampling function in Fourier space $\tilde{\rho}(\textbf{k})$, and $\Delta_{N\mathcal{F}}(\textbf{k})$ denotes a pixel noise contribution. The sampling function relates the one dimensional Lyman-alpha forest skewers to the 3D field and is given by
\be
\rho(\textbf{r})= \frac{1}{N} \sum_{i}w_{i}\delta_{D}^{2}(\textbf{r}_{\bot} -\textbf{r}_{\bot i})
\ee
with a set of weight functions $w_i$ that are used to maximize the SNR and $N$ is used to normalize $\int  \rho(\textbf{r}) d^3\textbf{r} = 1$. The sum extends over all the QSO locations $\textbf{r}_{\bot i} $ in the field.
The error in measurement of the cross power spectrum is hence given by $\delta P_{\mathcal{F}T}(\textbf{k})^2 =  \sigma_{\mathcal{F}T}^2/N_m$, where
\begin{equation}
\begin{split}
\sigma_{\mathcal{F}T}^2=& \frac{1}{2}\left[P_{\mathcal{F}T}^{2}(\textbf{k})+ \left(P_{\mathcal{FF}}(\textbf{k}) +P_{\mathcal{FF}}^{1D}(k_{\|}) P_{w}^{2D}+N_{\mathcal{F}}\right) \right. \\ & \left.\left( P_{TT}(\textbf{k}) + N_{T}\right)\right]
\end{split}
\end{equation}
where  $ P_{\mathcal{FF}}^{1D}(k_{\|})$ denotes the one dimensional power spectrum of Lyman-alpha forest flux fluctuations 
\begin{equation}
P_{\mathcal{FF}}^{1D}(k_{\|})=\frac{1}{(2\pi)^{2}}\int d^{2}\textbf{k}_{\bot}P_{\mathcal{FF}}(\textbf{k}).
\end{equation}
Here $N_T$ and $N_{\mathcal{F}}$ denotes the noise power spectrum for 21 cm observation and the Lyman-alpha forest respectively and $P_{w}^{2D}$ denotes the power spectrum of the weight functions.
The quantity $N_m$ is the number of available Fourier modes and is given by 
\begin{equation}
N_{m}(k,\mu)=\frac{2\pi k^{2} \mathcal{V}  dk d\mu}{(2\pi)^{3}}.
\end{equation}

The noise power spectrum for the 21 cm observations and the Lyman-alpha forest depend on the specific observational parameters. 
For radio-interferometric observation of the 21 cm signal the noise power spectrum is a function of the mode $k$ being probed and the observation frequency $\nu = 1420 ~{\rm MHz}/(1 + z)$ and is given by \citep{fg2, param1} 
\begin{equation}
N_{T}(k,\nu)=\frac{T_{sys}^{2}}{t_{0}}\left( \frac{c^{2}}{ \nu^2 A_{e}}\right) ^{2} \frac{2 r^{2}}{N_{ant}(N_{ant} -1) f_{2D}(U,\nu)} \frac{dr}{d\nu}.
\end{equation}
where, $A_e$ denotes the effective collecting area of a single  antenna, $ {\bf U} ={\bf k_{\perp}} r/2\pi $
and $t_o$ is the total observation time for a single pointing. The system temperature is denoted by $T_{sys}$. If $N_{ant}$ denotes the total number of antennae and  $f_{2D}(U, \nu) $ \citep{datta1, geilgrensler, navarro2} is the normalized baseline distribution function. 

For the Lyman-alpha forest observations, the noise power spectrum is given by \citep{mcquinnwhite}
\begin{equation}
N_{\mathcal{F}}=\sigma^{2}_{\mathcal{F} N}/\bar{n}
\end{equation}
where $\bar{n}$ is the average quasar number density in the field given by $\bar{n} = N_Q/ r^2 \theta_a^2$ for a total of $N_Q$ QSO spectra in the field of view and $\sigma^{2}_{\mathcal{F} N}$ is the variance of the pixel noise with 
\begin{equation}
\sigma^{2}_{\mathcal{F} N} = \frac{1}{N_Q} \sum {\bar{\mathcal{F}}}^{-2} [ S/N]_{\Delta x} ^{-2}\left( \frac{\Delta x}{1 \rm  Mpc} \right).
\end{equation}
 The signal to ratio $S/N$ in a pixel of size $\Delta x$, with $\Delta x$  matched to the frequency smoothing scale of the 21 cm observation is considered to be held at a constant average value of $3$.
To make error estimates for the two parameters $\Omega_{\nu}$ and $\Omega_{m}$ we  use the 
$2 \times 2$ Fisher matrix with elements 
\begin{equation}
F_{rs}=\int\int\frac{1}{\sigma_{{\mathcal F} T} ^2} \left( \frac{\partial P_{\mathcal{F}T}}{\partial\lambda_{r}}\right) \left( \frac{\partial P_{\mathcal{F}T}}{\partial\lambda_{s}}\right) \frac{2\pi k^{2} \mathcal{V} dk d\mu}{(2\pi)^{3}}
\end{equation}
where $\lambda_r$ and $\lambda_s$ can take values  $\Omega_{\nu}$ and $\Omega_m$.
The Cramer-Rao bound gives a limit on the the theoretical errors on  the $i^{th} $ parameter as $\delta \lambda_i = \sqrt{ F^{-1}_{ii}} $ which we use in this paper.

\section{Results and Discussion}
\label{sec:discussions}

We have assumed a fiducial cosmological model with
$(\Omega_{\Lambda},\Omega_{m},h, n_s, \sum\nolimits_{i}m_{i})$=$(0.685, 0.315, 0.6731, 0.9655, 0.1 {\rm eV})$  \citep{planck15} for this
analysis. The fiducial redshift $z = 2.5$ is chosen since the quasar
distribution peaks in the redshift range $ 2 < z < 3 $. At this
fiducial redshift the Lyman-alpha forest spectrum can be measured in
an approximate range $\Delta z = 0.43$. This is owing to the fact that
part of the quasar spectrum $ \sim 10,000 ~{\rm Km ~s} ^{-1}$ blue-ward
of the Lyman-alpha emission and $\sim 1000 ~{\rm Km ~s}^{-1}$ red-ward
of the Ly-$\beta$ line has to eliminated to avoid the quasar proximity
effect and contamination from Ly-$\beta$ and other metal lines
respectively. A high quasar density is required for the
cross-correlation to be measured at high level of sensitivity. We
consider a BOSS like Lyman-alpha forest survey with a quasar density
of $30 ~{\rm deg}^{-2}$ with an average $3\sigma$ sensitivity for the
measured spectra.  The BOSS survey covers a large part of the sky with
a transverse coverage of $ 10, 000  {\rm deg}^2$. The volume probed by
the Lyman-alpha forest is hence given by $ \frac{c}{H} \Delta z \times
r^2 \theta^2 $ with $\Delta z = 0.43 $ and $ \theta^2 = 3.043~ {\rm
  rad}^2$.  We note, however that the cross-correlation can only be
computed only in the overlapping volume between the Lyman-alpha forest
survey and 21cm survey volume which is expected to be much smaller.

We consider an 21cm intensity mapping experiment using a radio
interferometer with specifications roughly following that of
SKA1-mid \footnote{http://www.skatelescope.org/wp-content/uploads/2012/07/SKA-TEL-SKO-DD-001-1\_BaselineDesign1.pdf}. The radio interferometer shall operate in frequency range from $350$ MHz
to $14$ GHz.  The fiducial redshift of $ z = 2.5$ corresponds to a
frequency of $\sim 406\rm MHz$ falls in this frequency range.  We
assume the array to be composed of a total of $250$ dish like
antennae. Each individual antenna is of diameter $\sim 15$m with an
antenna efficiency of $0.8$.  We assume that the distribution of the
antennae is mostly condensed to a central core with $40\%$, $55\%$,
$70\%$, and $100\%$ of the total antennae are within $0.35$ km, $1$
km, $2.5$ km, and $100$ km radius respectively with density of
antennae falling off radially as an inverse square power law
\citep{navarro}.  This allows us to compute the normalized baseline
distribution function. It is evident that there can not be any
baselines below $30\rm{m}$. Typical intensity mapping experiments aim
towards higher power spectrum sensitivity at large scales.  The
telescope design supports this, since the telescopes are mostly
distributed in a central region. The baseline coverage at very small
non-linear scales is poor. We assume the system temperature $T_{sys}$
to be $30$K at observational frequency of $405.7$MHz. Further, the
telescope is assumed to have a $32 {\rm MHz}$ band width.

\begin{figure}\begin{center}
  \includegraphics[height=5cm, width=6.5cm, angle=0 ]{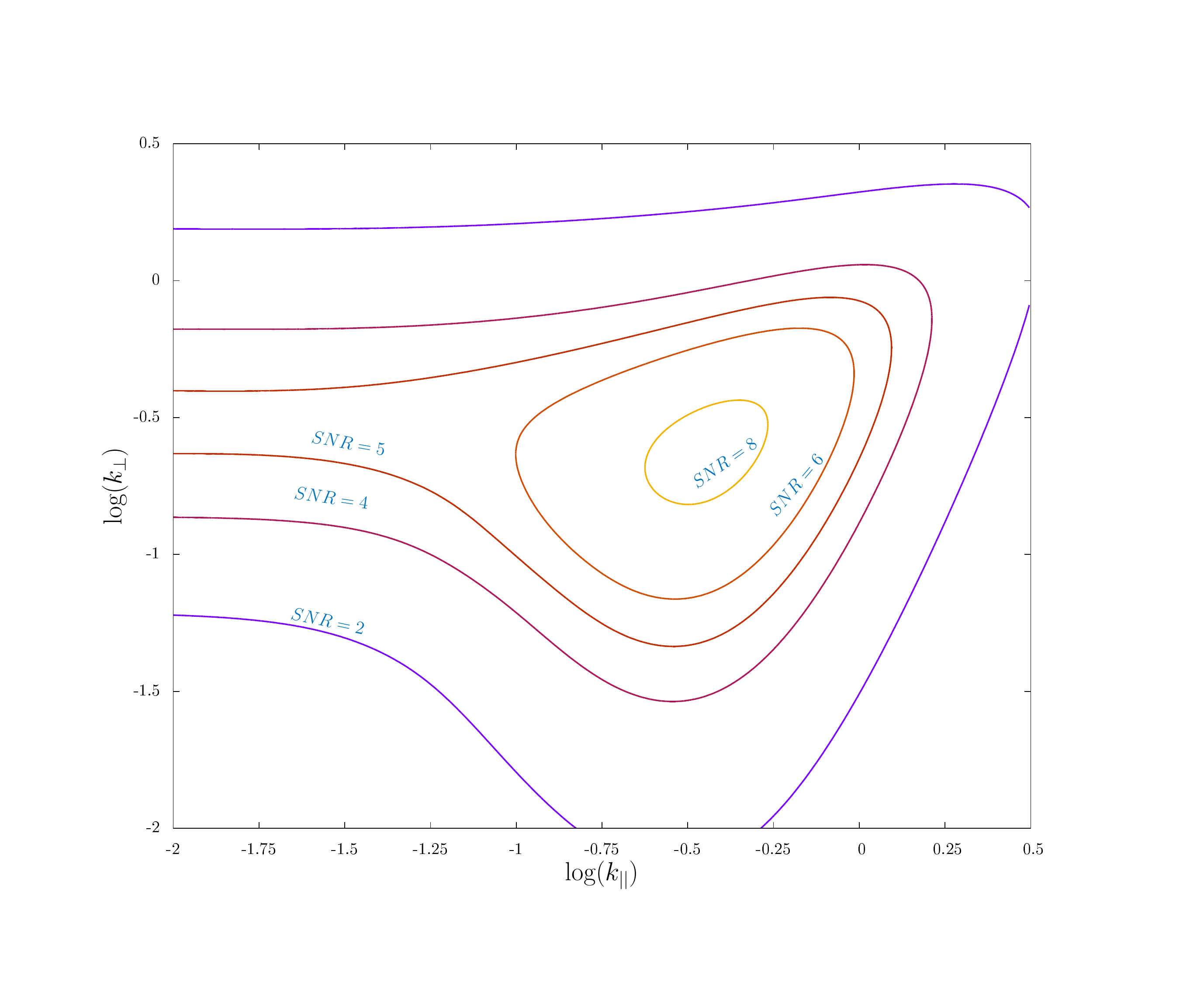}   \caption{ Signal to noise ratio for the cross-correlation power spectrum  for a single pointing radio observation of 400 hrs. We have considered a QSO  density 30 ${\rm deg}^{-2}$.}
\label{fig:snr}\end{center}
\end{figure}

Figure (\ref{fig:snr}) shows the signal to noise ratio for the
cross-correlation signal in the $(k_{\perp}, k_{\parallel})$ plane. We
have considered a $400 ~\rm hrs$ observation of the 21-cm signal in a
single pointing of the telescope. We find that a statistically
significant detection of the cross-correlation signal at a SNR $ > 8 $
is possible in the range $ 0.19 {\rm Mpc}^{-1} < k_{\parallel} <
0.6{\rm Mpc}^{-1} $ and $ 0.18{\rm Mpc}^{-1} < k_{\perp} < 0.36{\rm
  Mpc}^{-1} $. The anisotropy in the $k-$space owes its origin to the
redshift space distortion effects and the preferential sensitivity of
the noise in the $k-$space. We have assumed here that a perfect
foreground subtraction has been achieved. Foreground residuals are
expected to degrade the SNR. The issue of foreground removal has been
discussed later. The low sensitivity at small baselines owes it origin
to cosmic variance whereas the sensitivity at large scales is dictated
by instrument noise.

\begin{figure}
\begin{center}
  \includegraphics[height=5cm, width=5cm, angle=0]{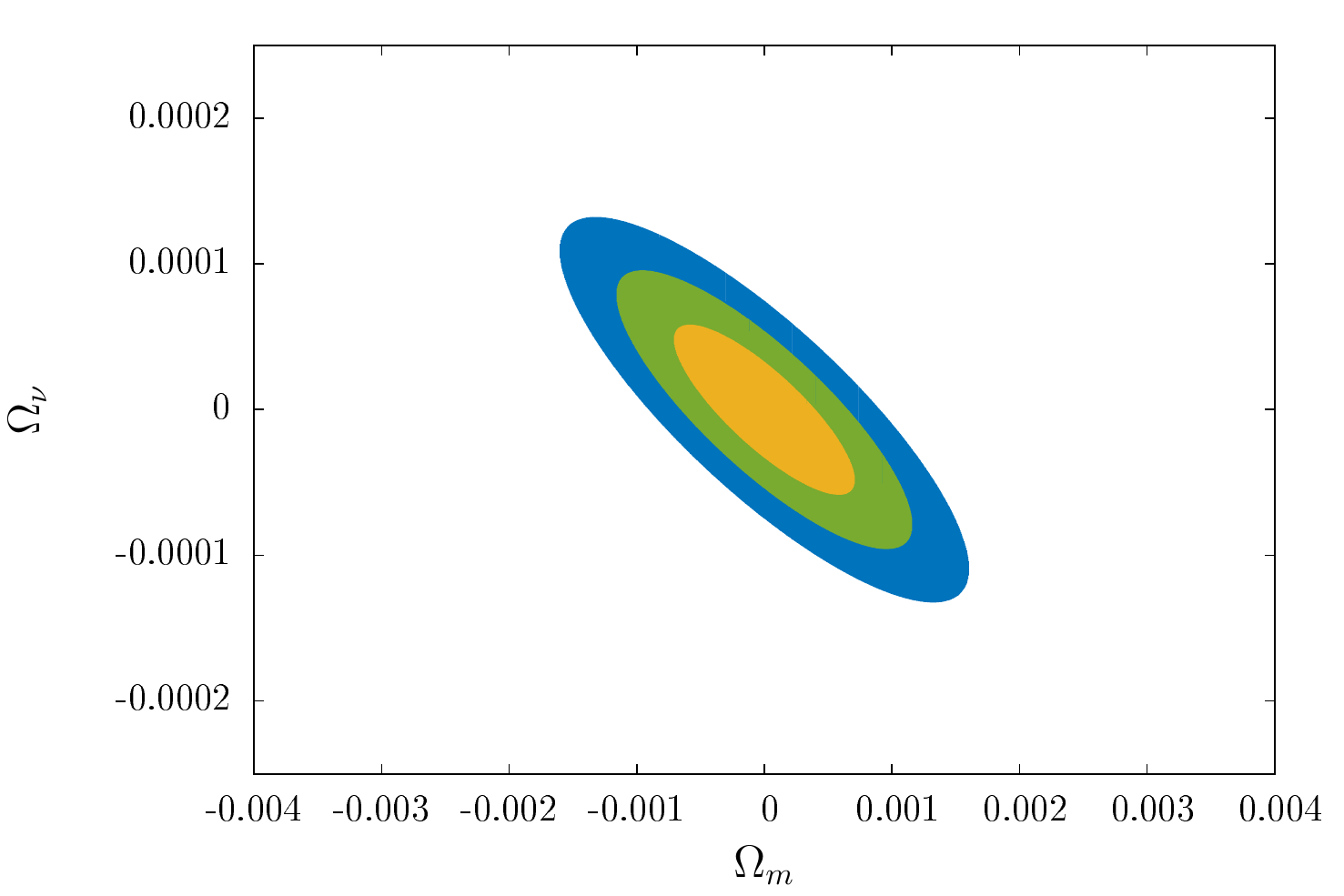}\\
  \includegraphics[height=3.5cm, width=3.5cm, angle=0]{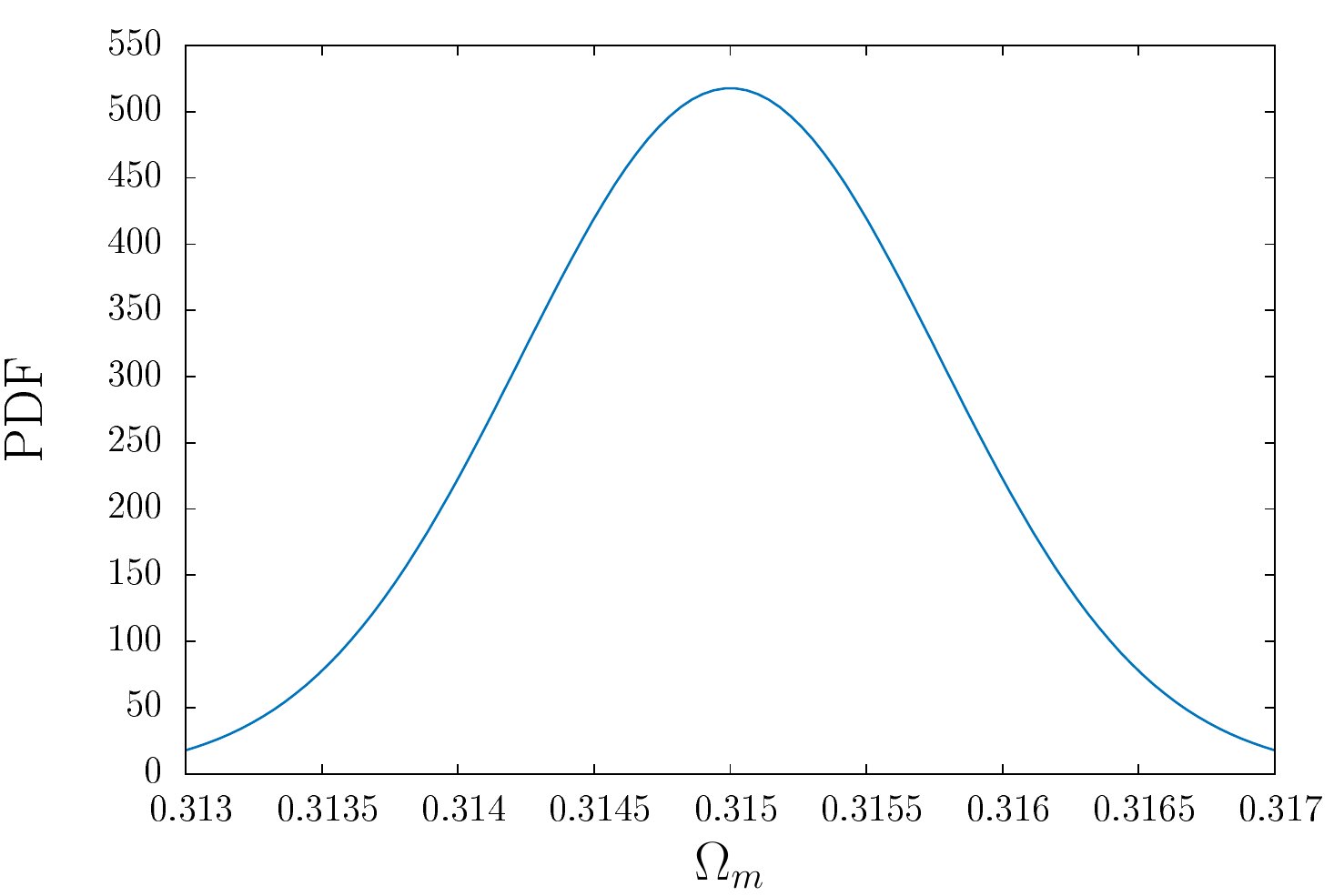}
  \includegraphics[height=3.5cm, width=3.5cm, angle=0]{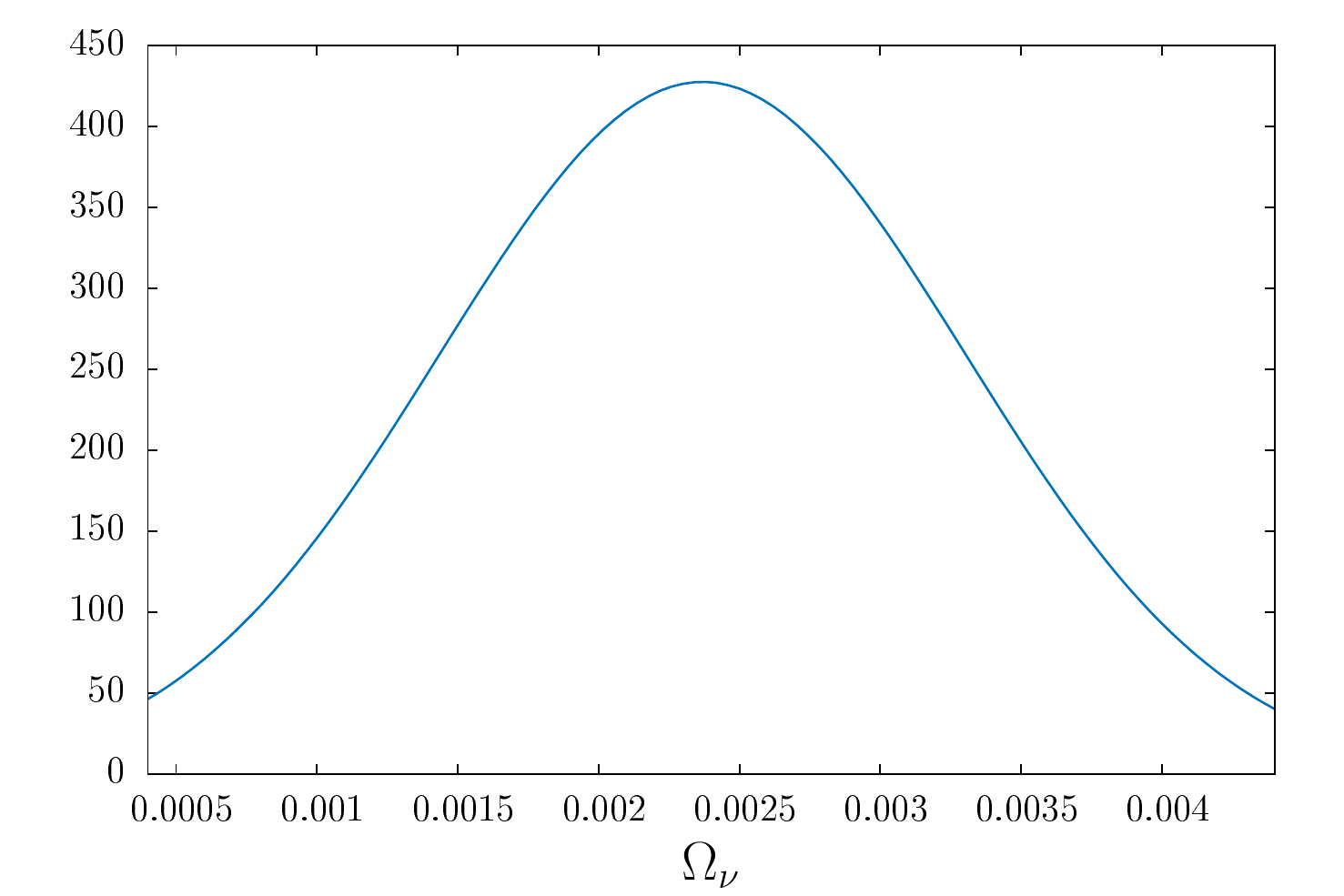}
    \caption{68.3\%,95.4\% and 99.8\% ellipse for 10000 hrs. observations for 25 pointings with each pointing of 400 hrs. The  marginalized one dimensional probability distribution function (PDF)
for $ \Omega_{m} $ and $ \Omega_{\nu} $ are also shown.}
\label{fig:ellipse}
\end{center}
\end{figure}

We next, look at the results of the Fisher matrix analysis towards
constraining the parameters $(\Omega_m, \Omega_{\nu} )$. Figure
(\ref{fig:ellipse}) shows the confidence ellipses at $68.3 \%$,
$95.4\%$ and $99.8\%$ levels. We have considered a total $10,000~ {\rm
  hrs} $ 21 cm observation for 25 pointings of radio telescope each
corresponding to a $400 ~{\rm hrs}$ observation. The Lyman-alpha
forest survey is assumed to have QSO number density of $30 {\rm  deg}^{-2}$ where each spectrum is measured at an average $> 3-\sigma$ pixel noise level. We find that for these observational parameters, $\Omega_m$ and $\Omega_{\nu}$ can be
measured at $0.29 \%$ and $3.25 \%$ respectively. Figure
(\ref{fig:ellipse}) also shows the marginalized probability
distribution for the parameters $\Omega_m$ and $\Omega_{\nu}$.
The SKA1-mid kind of radio interferometer considered here has a angular coverage of $2.8 \times 2.8 {\rm deg}^2$ at redshift of $2.5$.
This is much smaller than  volume covered by the  Lyman-alpha survey. The BOSS Lyman-alpha survey covers a much larger volume than a typical  21 cm observation.
A significant part of the Lyman-alpha survey volume can be used in the cross-correlation by considering multiple pointings of the radio interferometer.
We keep the total observation time fixed at $ 10,000 ~{\rm hrs}$ but now consider $50$ radio pointings each of $200 ~{\rm hrs}$ duration. Further we consider the possibility of dividing the total observing time into 100 pointings.The results show a clear improvement in the constraints.  We tabulate all the results in tables  (\ref{Tab:1}) and ( \ref{Tab:2}). This implies that for a QSO survey with a number density $ {\bar n} > 30 {\rm deg}^{-2}$ and with SKA like instrument, it is more advantageous to distribute the total observation time over as many fields of views as possible instead of deep imaging of a single radio field.

\begin{table}
\begin{center}
\begin{tabular}{|| c c c c c ||}
\hline 
\hline
Parameter  & Fiducial &  $\%$ Error & $\%$ Error& $\%$ Error \\
&value &25 pointings & 50 pointings &  100 pointings\\
\hline 
$ \Omega_{m} $ & 0.315 & .361 & 0.280 & 0.222 \\
\hline
$ \Omega_{m} $ & 0.315 & 0.310 & 0.237 & 0.184 \\
\hline
$ \Omega_{m} $ & 0.315 & 0.298 & 0.227 & 0.176 \\
\hline
$ \Omega_{m} $ & 0.315 & 0.296 & 0.227 & 0.174 \\
\hline
$ \Omega_{m} $ & 0.315 & 0.292 & 0.222 & 0.171 \\
\hline
\hline
\end{tabular}
\end{center}
\caption{1 $ \sigma $ error for various number of pointings for $ \Omega_{m} $ with a total 10,000hrs observation of the 21 cm signal}\label{Tab:1}
\end{table}

\begin{table}
\begin{center}
\begin{tabular}{|| c c c c c ||}
\hline 
\hline
Parameter & Fiducial  &  $\% $ Error & $  \% $ Error &  $\% $ Error  \\
 & value & 25 pointings& 50 pointings & 100 pointings\\
\hline 
$ \Omega_{\nu} $ & 7.10 $ \times  10^{-4}$ & 20.70 & 16.53 & 13.81 \\
\hline
$ \Omega_{\nu} $ & 1.42$ \times 10^{-3} $ & 6.47 & 5.21 & 4.25 \\
\hline
$ \Omega_{\nu} $ & 2.13 $\times 10^{-3} $ & 3.73 & 3.00 & 2.45 \\
\hline
$ \Omega_{\nu} $ & 2.37 $\times 10^{-3} $ & 3.25 & 2.61 & 2.13 \\
\hline
$ \Omega_{\nu} $ & 2.84 $\times 10^{-3} $ & 2.57 & 2.07 & 1.68 \\
\hline\hline
\end{tabular}
\end{center}
\caption{1-$\sigma $ error for various number of pointings for $ \Omega_{\nu} $ with a total 10,000 hrs observation of the 21 cm signal.}\label{Tab:2}
 \end{table}

 Up-to this point our analysis is focused on the estimates of the
 parameters HI 21 cm intensity mapping using (with SKA1-mid) and
 Lyman-alpha (BOSS) surveys, with some specific observational
 parameters  (i.e, QSO density, number of antennae,and time of
 observations).  In the next analysis our interest is focused on the
 variation of the error estimates if the above mentioned observational
 parameters are  changed  within the known possible range.
\begin{figure}
\begin{center}
  \includegraphics[height=3.5cm, width=6cm, angle=0]{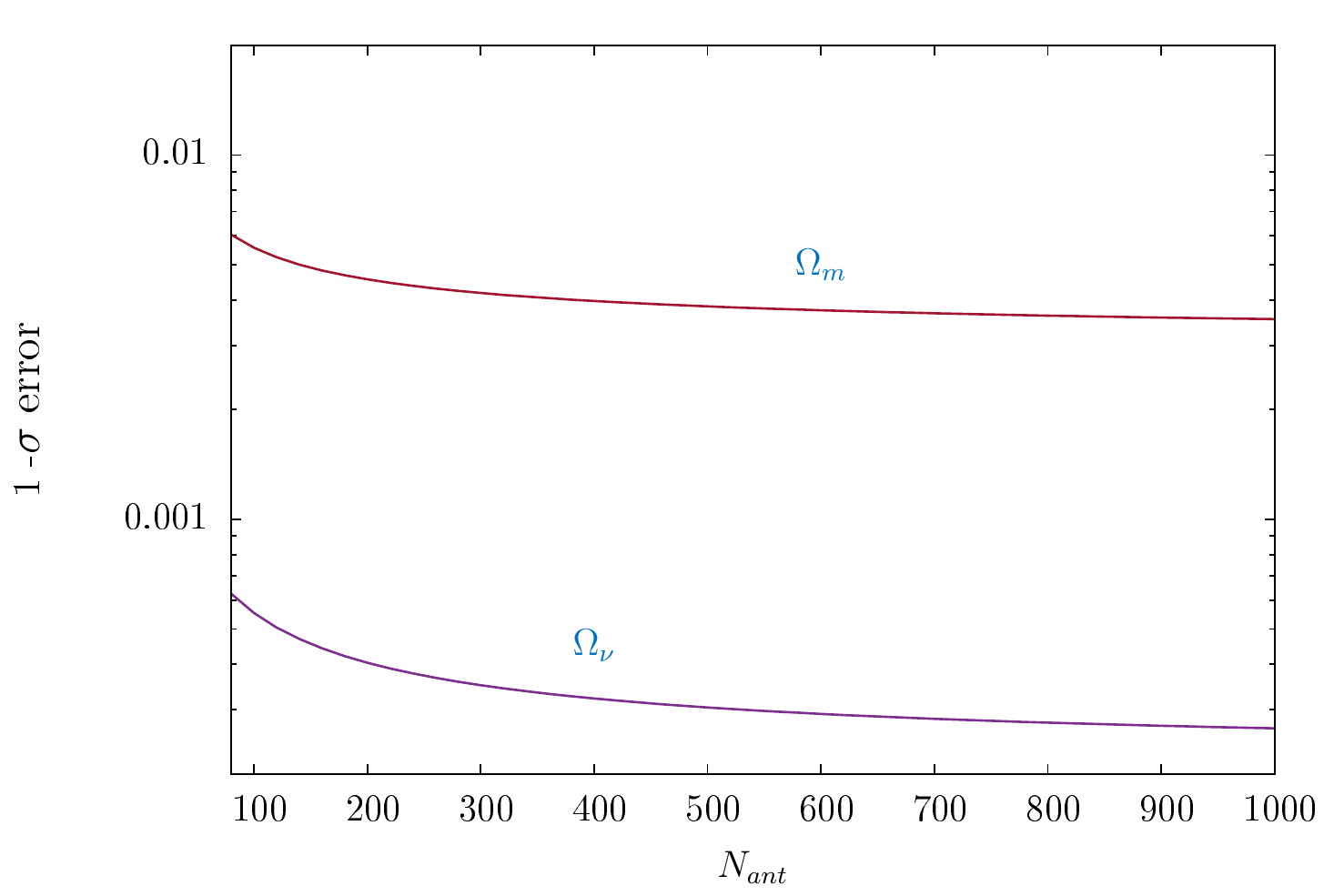}
  \caption{Variation of 1$-\sigma$ error of the parameters $\Omega_m$ and $\Omega_{\nu}$  with $ N_{ant} $ the number of antennas in the radio array.}
\end{center}
\label{fig:ante}
\end{figure}
  We first consider the effect of the variation of number of antennae
  from 100 to 1000 keeping the time of observation $ (t_{0}=400 {\rm
    hrs}) $ and QSO density ($ \bar{n}=30 {\rm deg}^{-2} $ )
  unchanged.  The total 10000 hrs time is hence divided over 25 radio
  pointings. The collecting area and resolution of the radio
  telescopes will increase with the number of antennae (keeping the
  dishes identical) . Figure (\ref{fig:ante}) shows the $1-\sigma $
  error bound on $ \Omega_{m},\Omega_{\nu} $ varying with the number
  of antennas in the radio array. We find that there is no significant
  improvement in the constraints when number of antennas are increased
  beyond 350 and, the error bound saturates to a limit set by cosmic
  variance and the parameters of the Lyman-alpha survey.
\begin{figure}
\begin{center}
  \includegraphics[height=3.5cm, width=6.0cm, angle=0]{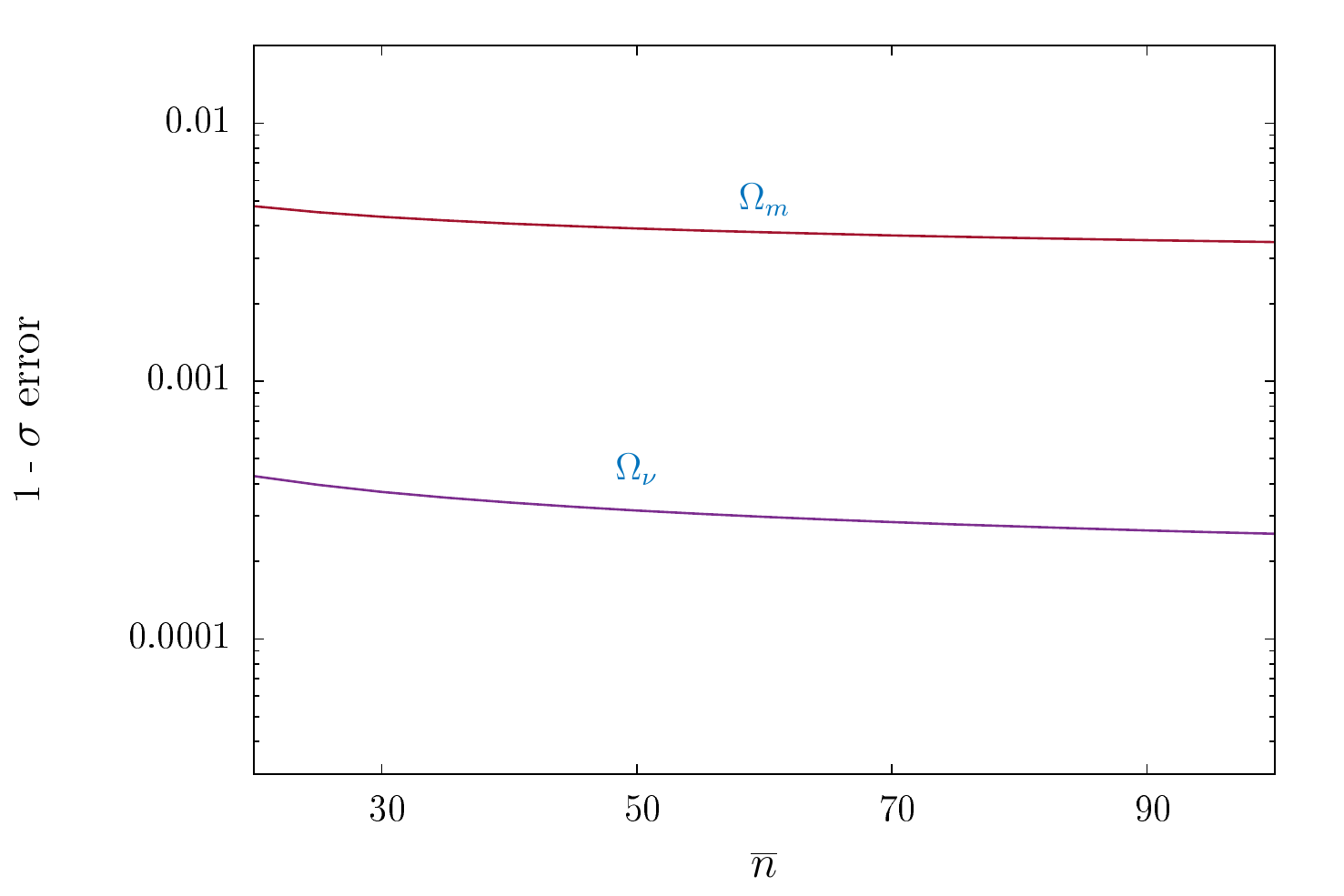}
  \caption{Variation of 1$-\sigma$ error of the parameters $\Omega_m$ and $\Omega_{\nu}$  with $ \bar{n} $.}
\end{center}
\label{fig:qsr}
\end{figure}
The noise contribution from the Lyman-alpha forest power spectrum
measurement depends on the number density of Quasars in the survey.
Figure (\ref{fig:qsr}) represents the improvement of 1$ \sigma $ error
in the estimation of parameters due to the of variation QSO density ($
\bar{n} $).  In this case we find that beyond $ \bar{n}> 50 $ the
decline of the error is very slow and the cosmic variance limit is
asymptotically reached.
\begin{figure}
\begin{center}
 \includegraphics[height=4cm, width=6cm, angle=0]{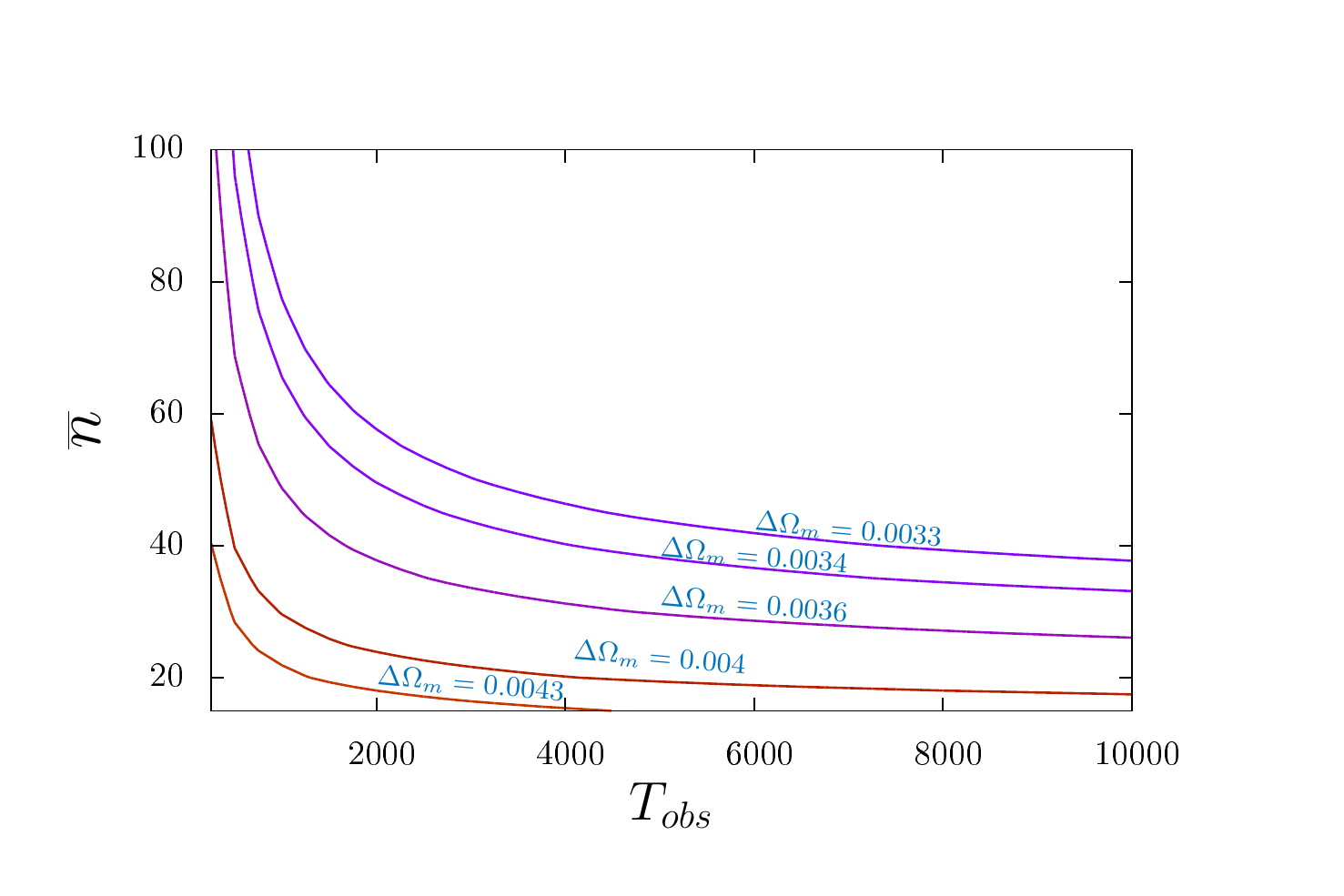}
 \includegraphics[height=4cm, width=6cm, angle=0]{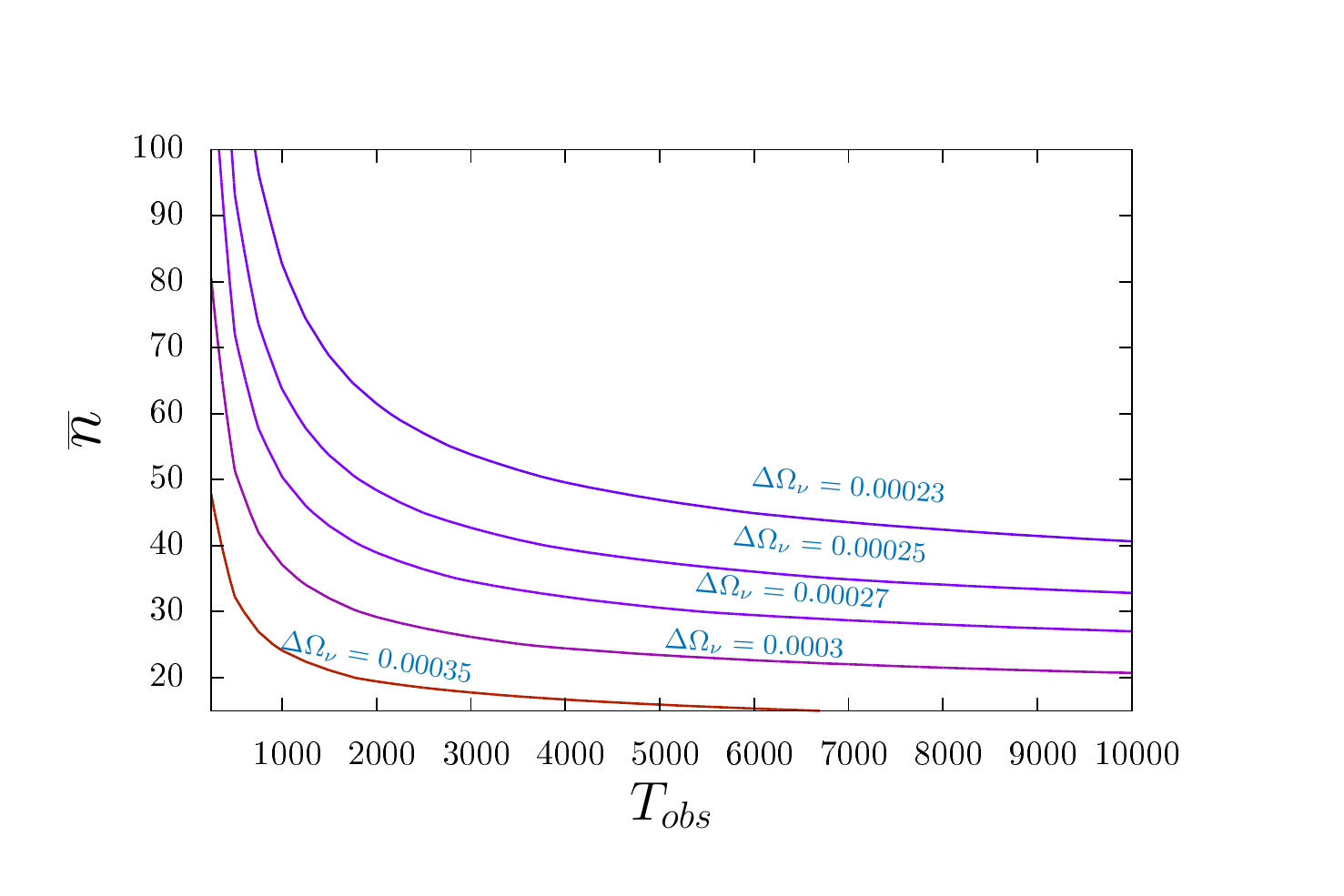}
 \caption{$\Omega_{m}$ and $\Omega_{\nu}$ contours in the  $\bar{n}$ and time of observation plane.}
\label{fig:om}
\end{center}
\end{figure}

For a more practical scenario it is useful to investigate the optimal
observational strategy. We characterize the radio observation with
the time of observation and Lyman-alpha forest survey with the quasar
number density in the survey with high SNR spectra. Figure
(\ref{fig:om}) shows the contours of constant 1 $ \sigma $ errors of $
\Omega_{m} , \Omega_{\nu} $ for a single pointing radio observation.
The bottom left corresponds to large observational noise owing to
small quasar number density and small time of observation. The top
right corner corresponds to large quasar density and long duration
radio observation.  The error in the parameters show a decrease as one
moves from the bottom left to the top right corner through the
contours. However, beyond a point there is no further improvement and
this corresponds to the cosmic variance limit. In this limit the
parameter $\Omega_m$ can be measured at a $ 0.9\% $ level and
$\Omega_{\nu}$ can be measured at $ 7.6 \%$. Further improvement is
only possible through increase of the number of pointings which
increases the survey volume and the errors scale as $\sim 1/\sqrt
N_{poin}$. A total of 25,000 hrs radio observation distributed equally
over 25 pointings and a Lyman-alpha survey with $\bar n = 60
{\rm deg} ^{-2}$ will allow $\Omega_{\nu}$ to be measured at a $ 2.26 \%$
level.  This corresponds to a measurement of $\sum m_{\nu}$ at the
precision of $(100 \pm 2.26) \rm meV$ and $f_{\nu}$ at $2.49 \%$. Our
general observation is that there should be greater emphasis of
increasing survey volumes through bigger bandwidths or using smaller
dishes and considering multiple fields of views against extremely deep
surveys in small fields. This shall make the constraints parameters
stronger. we however  note that 25,000 hrs observation in 25 pointings is a
rather unfeasible observational strategy. In this idealized survey it is possible to even look into the mass hierarchy for neutrinos.
However, Several observational issues poses severe problems towards detection
of the signal. The crucial issue as regarding the 21 cm signal, is the
issue of foregrounds. Foregrounds like the galactic synchrotron
radiation and extra galactic point sources are several orders of
magnitude larger than the cosmological 21 cm signal. Further, man made
RFIs \citep{rfi} also contaminate the signal severely. Several foreground
subtraction methods have been proposed. We note that though foreground
subtraction is crucial towards measuring the 21 cm signal, these
contaminants appear as noise in the cross-correlation and foreground
residuals even of the same order of magnitude as the signal shall not
entirely inhibit the detection of the cross-correlation signal
unlike the unlike the auto correlation. 
Assuming a fiducial model with $\sum m_{\nu} = 0.1 {\rm eV}$ we find that if we have a foreground residual of the 21-cm signal at  200\% of the signal itself then the constraint on $\Omega_{\nu}$ degrades from $2.26 \%$  to $ 2.7 \%$. Further, if the Lyman-$\alpha$ forest pixel noise is at $1-\sigma$, the constaint on neutrino mass degrades to $ \sim 6\%$.

The Lyman-alpha forest surveys
require very high SNR measurement of a large number of spectra for
high precision detection of the cross-correlation signal which is
necessary for obtaining stringent bounds on the neutrino
mass. Further, one has to worry about several other observational
issues like the effect of Galactic super wind. contamination from
other metal lines etc.  We conclude by noting that with future
radio-observations of the cosmological redshifted 21-cm signal and
Lyman-alpha forest surveys, it is possible to measure neutrino mass at
a high level of precision using the cross-correlation power spectrum.
We predict that the bounds obtained from such a measurement shall be
competitive with other probes.
\section{Acknowledgements}
TGS would like to acknowledge the Department of Science and Technology (DST), Government of India for providing financial support through the project SR/FTP/PS-172/2012. 

\bibliographystyle{mnras}
\bibliography{references}

\end{document}